\documentstyle[twocolumn,aps,psfig]{revtex}
\begin{document}
\draft
\flushbottom
\twocolumn[
\hsize\textwidth\columnwidth\hsize\csname @twocolumnfalse\endcsname

\title{ Surface plasmon toy-model of a rotating black hole.}
\author{Igor I. Smolyaninov}
\address{ Department of Electrical and Computer Engineering \\
University of Maryland, College Park,\\
MD 20742}
\date{\today}
\maketitle
\tightenlines
\widetext
\advance\leftskip by 57pt
\advance\rightskip by 57pt

\begin{abstract}
Recently introduced surface plasmon toy black hole model has been extended in order to emulate a rotating black hole (Kerr metric). Physical realization of this model involves a droplet of an optically active liquid on the metal surface which supports propagation of surface plasmons. Such droplets are shown to exhibit giant optical activity in the frequency range near the surface plasmon resonance of a metal-liquid interface.    
\end{abstract}

\pacs{PACS no.: 78.67.-n, 04.70.Bw }
]
\narrowtext

\tightenlines

The realization that solid-state toy models may help in an understanding of electromagnetic phenomena in curved space-time has led to considerable recent effort in developing toy models of electromagnetic \cite{1} and sonic \cite{2} black holes. In the case of media electrodynamics this is possible because of an analogy between the propagation of light in matter and in curved space-times: it is well known that the Maxwell equations in a general curved space-time background $g_{ik}(x,t)$ are equivalent to the phenomenological Maxwell equations in the presence of a matter background with nontrivial electric and magnetic permeability tensors $\epsilon _{ij}(x,t)$ and $\mu _{ij}(x,t)$ \cite{3}. In this analogy, the event horizon corresponds to a surface of singular electric and magnetic permeabilities, so that the speed of light goes to zero, and light is "frozen" near such a surface. In the absence of established quantum gravitation theory the toy models are helpful in understanding electromagnetic phenomena in curved space-times, such as Hawking radiation \cite{4} and the Unruh effect \cite{5}. Unfortunately, until recently all the suggested electromagnetic black hole toy-models were very difficult to realize and study experimentally, so virtually no experimental work was done in this field. Very recently a simple surface plasmon toy black hole model has been suggested and observed in the experiment \cite{6}. This model is based on the surface plasmon resonance at the edge of a liquid droplet deposited on the metal surface, which supports propagation of surface plasmons. Surface plasmons are collective excitations of the conductive electrons and the electromagnetic field \cite{7}. They exist in "curved three-dimensional space-times" defined by the shape of the metal-dielectric interface. Since in many experimental geometries surface plasmons are weakly coupled to the outside world (to free-space photons, phonons, single-electron excitations, etc.) it is reasonable to treat the physics of surface plasmons separately from the rest of the surface and bulk excitations, so that a field-theory of surface plasmons in a curved space-time background may be considered. For example, a nanohole in a thin metal membrane may be treated as a "wormhole" connecting two "flat" surface plasmon worlds located on the opposite interfaces of the membrane \cite{8}. On the other hand, near the plasmon resonance (which is defined by the condition that $\epsilon _m(\omega )=-\epsilon _d$ at the metal-dielectric interface, where $\epsilon _m(\omega )$ and $\epsilon _d$ are the dielectric constants of metal and dielectric, respectively \cite{7}) the surface plasmon velocity vanishes, so that the surface plasmon "stops" on the metal surface, and the surface charge and the normal component of the electric field diverge. For a given frequency of light, the spatial boundary of the plasmon resonance ("the event horizon" of our toy model) may be defined at will using the geometry $\epsilon _d(x,y)$ of the absorbed layer of dielectric on the metal surface. Thus, the plasmon resonance becomes a natural candidate to emulate the event horizon of a black hole. As a result, toy two-dimensional surface plasmon black holes can be easily produced and studied \cite{6}. 

In what follows I am going to extend the surface plasmon toy black hole model in order to emulate the space-time metric of a rotating black hole (Kerr metric \cite{9}). Physical realization of this model involves a droplet of an optically active liquid on the metal surface which supports propagation of surface plasmons. I will show that the droplets of such liquids on the metal surface exhibit giant optical activity in the frequency range near the surface plasmon resonance of a metal-liquid interface.    

As a first step, let us consider in detail the dispersion law of a surface plasmon (SP), which propagates along the metal-dielectric interface. The SP field decays exponentially both inside the metal and the dielectric. Inside the dielectric the decay exponent is roughly equal to the SP wave vector. As a first step let us assume that both metal and dielectric completely fill the respective $z<0$ and $z>0$ half-spaces. In such a case the dispersion law can be written as \cite{7} 

\begin{equation}  
k^2=\frac{\omega ^2}{c^2}\frac{\epsilon _d\epsilon _m(\omega )}{\epsilon _d+\epsilon _m(\omega)} ,
\end{equation}

where we will assume that $\epsilon _m=1-\omega _p^2/\omega ^2$ according to the Drude model, and $\omega _p$ is the plasma frequency of the metal. This dispersion law is shown in Fig.1(b) for the cases of metal-vacuum and metal-dielectric interfaces. It starts as a "light line" in the respective dielectric at low frequencies and approaches asymptotically $\omega =\omega _p/(1+\epsilon _d)^{1/2}$ at very large wave vectors. The latter frequency corresponds to the so-called surface plasmon resonance. Under the surface plasmon resonance conditions both phase and group velocity of the SPs is zero, and the surface charge and the normal component of the electric field diverge. Since at every wavevector the SP dispersion law is located to the right of the "light line", the SPs of the plane metal-dielectric interface are decoupled from the free-space photons due to the momentum conservation law.   

If a droplet of dielectric (Fig.1(a)) is placed on the metal surface, the SP dispersion law will be a function of the local thickness of the droplet. Deep inside the droplet far from its edges the SP dispersion law will look similar to the case of a metal-dielectric interface, whereas near the edges (where the dielectric is thinner) it will approach the SP dispersion law for the metal-vacuum interface. As a result, for every frequency between $\omega _p/(1+\epsilon _d)^{1/2}$ and $\omega _p/2^{1/2}$ there will be a closed linear boundary inside the droplet for which the surface plasmon resonance conditions are satisfied. Let us show that such a droplet of dielectric on the metal interface behaves as a "surface plasmon black hole" in the frequency range between $\omega _p/(1+\epsilon _d)^{1/2}$ and $\omega _p/2^{1/2}$, and that the described boundary of the surface plasmon resonance behaves as an "event horizon" of such a black hole. 

Let us consider a SP within this frequency range, which is trapped near its respective "event horizon", and which is trying to leave a large droplet of dielectric (see Fig.2). The fact that the droplet is large means that ray optics may be used. Since the component of the SP momentum parallel to the droplet boundary has to be conserved, such a SP will be totally internally reflected by the surface plasmon resonance boundary back inside the droplet at any non-zero angle of incidence. This is a simple consequence of the fact that near the "event horizon" the effective refractive index of the droplet for surface plasmons is infinite (according to eq.(1), both phase and group velocity of surface plasmons is zero at surface plasmon resonance). On the other hand, even if the angle of incidence is zero, it will take the SP infinite time to leave the resonance boundary. Thus, the droplet behaves as a black hole for surface plasmons, and the line near the droplet boundary where the surface plasmon resonance conditions are satisfied plays the role of the event horizon for surface plasmons. 

The toy black holes described above are extremely easy to make and observe \cite{6}. In our experiments a small droplet of glycerin was placed on the gold film surface and further smeared over the surface using lens paper, so that a large number of glycerin microdroplets were formed on the surface (Fig.3(a)). These microdroplets were illuminated with white light through the glass prism (Fig.1(a)) in the so-called Kretschman geometry \cite{7}. The Kretschman geometry allows for efficient SP excitation on the gold-vacuum interface due to phase matching between the SPs and photons in the glass prism. As a result, SPs were launched into the gold film area around the droplet. Photograph taken under a microscope of one of such microdroplets is shown in Fig.3(b). The white rim of light near the edge of the droplet is clearly seen. It corresponds to the effective SP event horizon described above. Near this toy event horizon SPs are stopped or reflected back inside the droplet. In addition, a small portion of the SP field may be scattered out of the two-dimensional surface plasmon world into normal three-dimensional photons. These photons produced the image in Fig.3(b). We also conducted near-field optical measurements of the local surface plasmon field distribution around the droplet boundary Fig.3(c,d,e) using a sharp tapered optical fiber as a microscope tip. These measurements were performed similar to the measurements of surface plasmon scattering by individual surface defects described in \cite{10}. The droplet was illuminated with 488 nm laser light in the Kretschman geometry. The tip of the microscope was able to penetrate inside the glycerin droplet, and measure the local plasmon field distribution both inside and outside the droplet. Inside the droplet (in the right half of the images) the shear-force image (c) corresponds to the increase in viscous friction rather than the droplet topography. However, this image accurately represents the location of the droplet boundary, shown by the arrow in Fig.3(e). The sharp and narrow local maximum of the surface plasmon field just inside the droplet near its boundary is clearly visible in the near-field image Fig.3(d) and its cross-section Fig.3(e).         

Unfortunately, the described toy SP black hole model does not work outside the frequency range between $\omega _p/(1+\epsilon _d)^{1/2}$ and $\omega _p/2^{1/2}$. On the other hand, this is a common feature of every electromagnetic toy black hole model suggested so far. All such toy models necessarily work only within a limited frequency range. In addition, the losses in metal and dielectric described by the so far neglected imaginary parts of their dielectric functions will put a stop to the singularities of the field somewhere very near the toy event horizon (however, such "good" metals as gold and silver have very small imaginary parts of their dielectric constants, and hence, very pronounced plasmon resonances \cite{7}). Notwithstanding these limitations, the ease of making and observing such toy SP black holes makes them a very promising research object. If we forget about the language of "black holes" and "event horizons" for a moment, the SP optics phenomenon represented in Fig.3(b) remains a potentially very interesting effect in surface plasmon optics. Namely, this photo shows the existence of a two-dimensional SP analog of whispering gallery modes, which are well-known in the optics of light in droplets and other spherical dielectric particles. Whispering gallery modes in liquid microdroplets are known to substantially enhance nonlinear optical phenomena due to cavity quantum electrodynamic effects \cite{11}. One may expect even higher enhancement of nonlinear optical mixing in liquid droplets on the metal surfaces due to enhancement of surface electromagnetic field inherent to surface plasmon excitation, and in addition, due to accumulation of SP energy near the surface plasmon event horizons at the droplet boundaries. This strong enhancement of nonlinear optical effects in liquid droplets may be very useful in chemical and biological sensing applications. 

Let us now try and extend this surface plasmon toy black hole model so that a rotating black hole metric (Kerr metric) can be emulated. The relationship between the $\vec{D}$ and $\vec{E}$ fields in a gravitational field with the metric $g_{ik}$ can be written in the form \cite{9}:

\begin{equation}  
\vec{D}=\frac{\vec{E}}{h^{1/2}}+\vec{H}\times \vec{g} ,
\end{equation}

where $h=g_{00}$ and $g_{\alpha }=-g_{0\alpha }/g_{00}$. For an electromagnetic wave this relationship is very similar to the relationship between the $\vec{D}$ and $\vec{E}$ fields in the optically active medium, if we identify $1/h^{1/2}$ as $\epsilon $ of the medium, and recall how optical activity is introduced in media electrodynamics.  

There are different ways of introducing optical activity (gyration) tensor in the macroscopic Maxwell equations. It can be introduced in a symmetric form, which is sometimes called Condon relations \cite{12}:

\begin{equation}  
\vec{D}=\epsilon \vec{E}+\gamma\frac{\partial \vec{B}}{\partial t}
\end{equation}

\begin{equation}  
\vec{H}=\mu ^{-1}\vec{B}+\gamma\frac{\partial \vec{E}}{\partial t}
\end{equation}

Or it can be introduced only in an equation for $\vec{D}$ (see \cite{13,14}). In 
our consideration I will follow Landau and Lifshitz \cite{13}, and for simplicity use only the following equation valid in isotropic or cubic-symmetry materials:

\begin{equation}  
\vec{D}=\epsilon \vec{E}+i\vec{E}\times \vec{g} ,
\end{equation}

where $\vec{g}$ is called the gyration vector. If the medium exhibits magneto-
optical effect and does not exhibit natural optical activity $\vec{g}$ is 
proportional to the magnetic field $\vec{H}$:

\begin{equation}  
\vec{g}=f\vec{H} ,
\end{equation}

where the constant $f$ may be either positive or negative. For metals in the 
Drude model at $\omega >>eH/mc$

\begin{equation}  
f(\omega )= -\frac{4\pi Ne^3}{cm^2\omega ^3}=-\frac{e\omega _p^2}{mc\omega ^3} ,
\end{equation}

where $\omega _p$ is the plasma frequency and $m$ is the electron mass \cite{13}. In any case, the local difference in refractive index for the left $n_-$ and right $n_+$ circular polarizations of light is proportional to the local value of the gyration vector $\vec{g}$.

Thus, from the relationships between the $\vec{D}$ and $\vec{E}$ fields in a gravitational field and in media electrodynamics we conclude that space-time behaves as a chiral optical medium if $\vec{g}\neq 0$ (we must take into account different conventions for the use of imaginary numbers implemented in \cite{9} and \cite{13}, and hence equations (2) and (5)).

A space-time region with $\vec{g}\neq 0$ can be described locally as a rotating coordinate frame with an angular velocity \cite{9}

\begin{equation}  
\vec{\Omega }=\frac{ch^{1/2}}{2}\vec{\nabla }\times \vec{g} 
\end{equation}
 
Similar vector $\vec{\Omega }$ field can be defined for any chiral medium regardless of the nature of its optical activity (natural or magnetic field induced). Thus, electrodynamics of such medium may be understood as if there is a distribution of local angular rotation $\vec{\Omega }$ field inside the medium.

Now it is clear that in order to emulate a rotating black hole we must use a droplet of liquid which exhibits optical activity. Let us consider a medium with $\vec{g}=\vec{g}(r)$ directed along the $\phi $-coordinate (Fig.4(b)). Such a distribution may be produced, for example, in a twisted layer of liquid crystal or in a liquid droplet exhibiting magneto-optical effect if a droplet is formed around a small cylindrical wire and a current is passed through the wire. Such a geometry will result in different refractive indices $n_+$ and $n_-$ of the liquid as seen by the surface plasmons, which propagate in the clockwise (right) and counterclockwise (left) directions near the droplet boundary. As a result, these plasmons will have slightly different dispersion laws (Fig.4(a)). This difference produces two main effects: the position of the effective event horizon for the left and right plasmons will be different inside the liquid droplet (Fig.4(b)); and there will be a narrow frequency range (located near the surface plasmon resonance of the metal-liquid interface) where the event horizon will exist for only one (left or right) kind of surface plasmons (Fig.4(a)). 

The first effect closely resembles the properties of the Kerr metric, which characterizes a rotating black hole. The Kerr metric has two characteristic surfaces around a rotating black hole. There exists a spherical event horizon similar to the case of a non-rotating black hole, and there is a so called ergosphere \cite{9} just outside the black hole event horizon, which corresponds to the area of space where every particle must rotate in the direction of the black hole rotation. There are no particle states, which rotate in the opposite direction inside the ergoshere. Similar to the case of a real rotating black hole, our toy surface plasmon black hole has a toy-ergosphere: between the two event horizons shown in Fig.4(b) there are no surface plasmon states which rotate in the direction opposite to the direction of the effective $\vec{\Omega }$ field of the toy black hole. This area of space is quite unusual from the point of view of surface plasmons. We may say that even the surface plasmon vacuum rotates in this area, similar to the description of the electromagnetic vacuum inside the band gap of a photonic crystal as being "colored" vacuum. Further study of the linear and nonlinear optical properties of such rotating vacuum may be quite interesting from the fundamental optics point of view.

The second effect may be quite important from the point of view of practical applications. In the frequency range where the effective event horizon exists only for one (left or right) kind of plasmons, the array of droplets will behave as a medium which exhibits giant planar optical activity. Such media are being actively developed at the moment \cite{15}.  

In conclusion, recently introduced surface plasmon toy black hole model has been extended in order to emulate a rotating black hole (Kerr metric). Physical realization of this model involves a droplet of an optically active liquid on the metal surface which supports propagation of surface plasmons. Such droplets are shown to exhibit giant optical activity in the frequency range near the surface plasmon resonance of a metal-liquid interface.   

This work has been supported in part by the NSF grants ECS-0210438 and ECS-0304046.

Figure captions.

Fig.1 (a) Experimental geometry of surface plasmon toy black hole observation. (b) Surface plasmon dispersion law for the cases of metal-vacuum and metal-dielectric interfaces

Fig.2 A surface plasmon trapped inside a droplet near the effective "event horizon": The projection of surface plasmon momentum parallel to the droplet edge must be conserved. Due to effectively infinite refractive index near the droplet edge surface plasmons experience total internal reflection at any angle of incidence. 

Fig.3 Far-field (a,b) and near-field (c,d) images of a toy surface plasmon black hole: (a) Droplet of glycerin on a gold film surface (illuminated from the top). The droplet diameter is approximately 15 micrometers. (b) The same droplet illuminated with white light in the Kretschman geometry, which provides efficient coupling of light to surface plasmons on the gold-vacuum interface (Fig.1(a)). The white rim around the droplet boundary corresponds to the effective surface plasmon "event horizon". (c) and (d) show $10\times 10 \mu m^2$ topographical and near-field optical images of a similar droplet boundary (droplet is located in the right half of the images) illuminated with 488 nm laser light. Cross-sections of both images are shown in (e). Position of a droplet boundary is indicated by the arrow.  

Fig.4 (a) Surface plasmon dispersion law for the metal-vacuum and metal-dielectric interfaces in the case of optically active dielectric. (b) Difference in refractive indices for the left and right surface plasmons results in different locations of the effective event horizons for these plasmons. The local directions of the gyration filed are shown by the arrows.

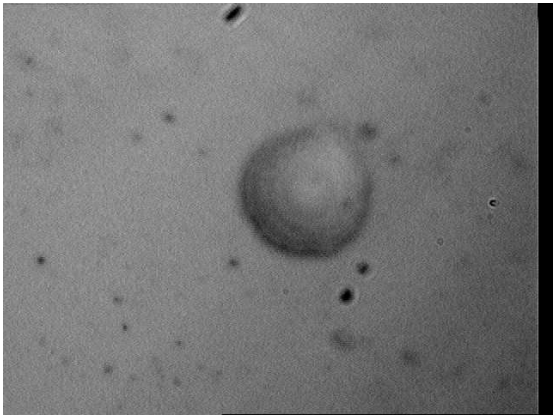
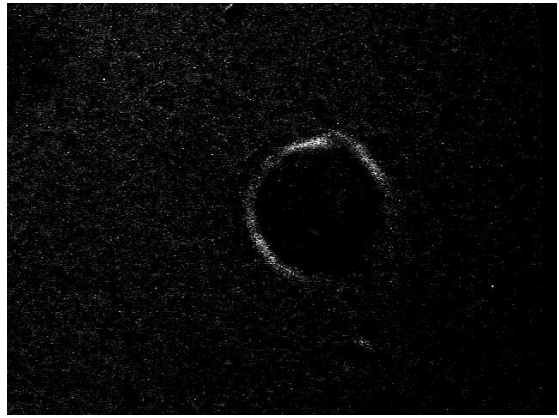
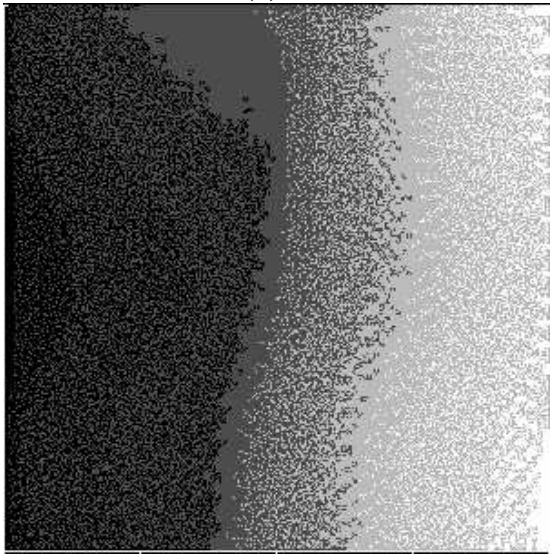
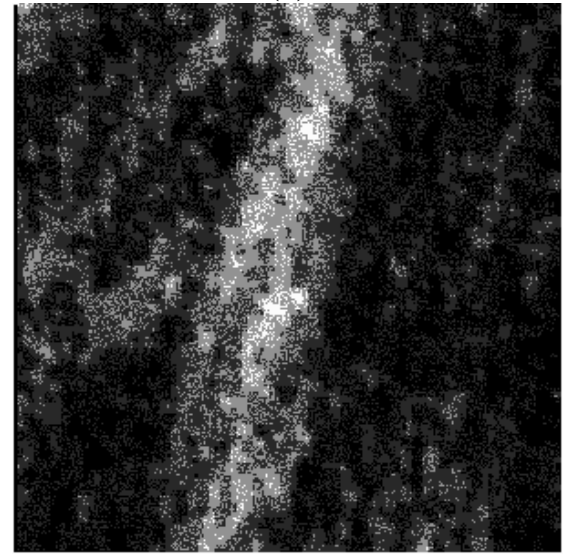
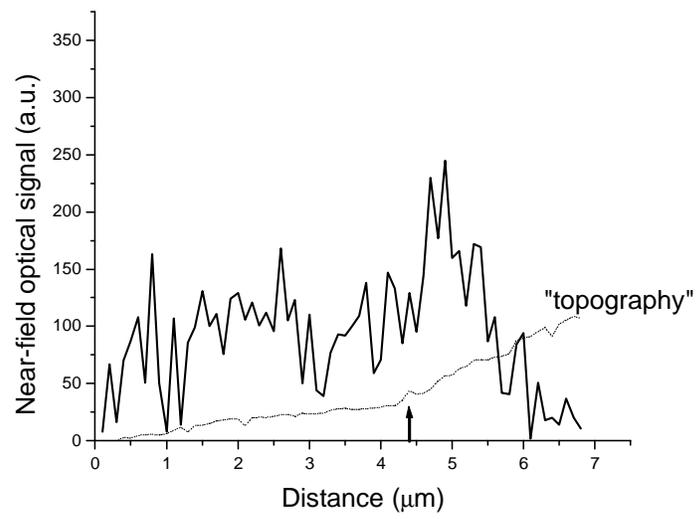
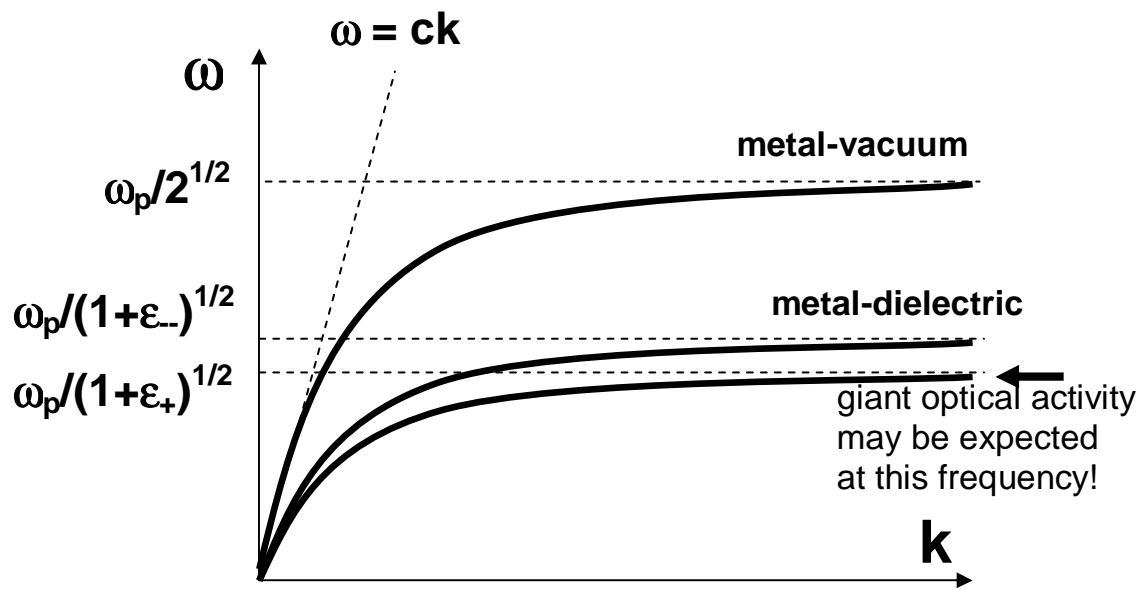
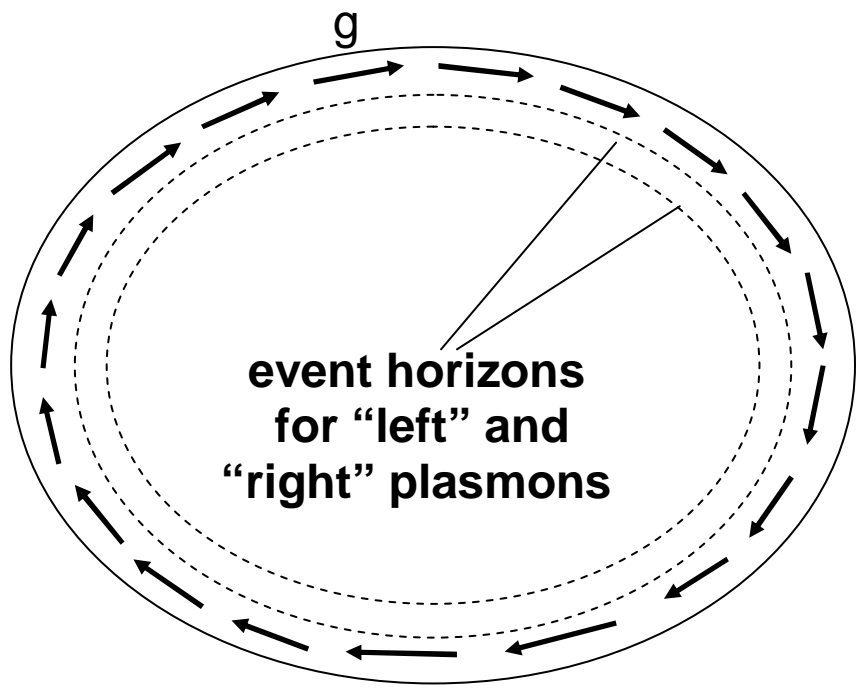
\end{document}